# Physics Archives

June 2008

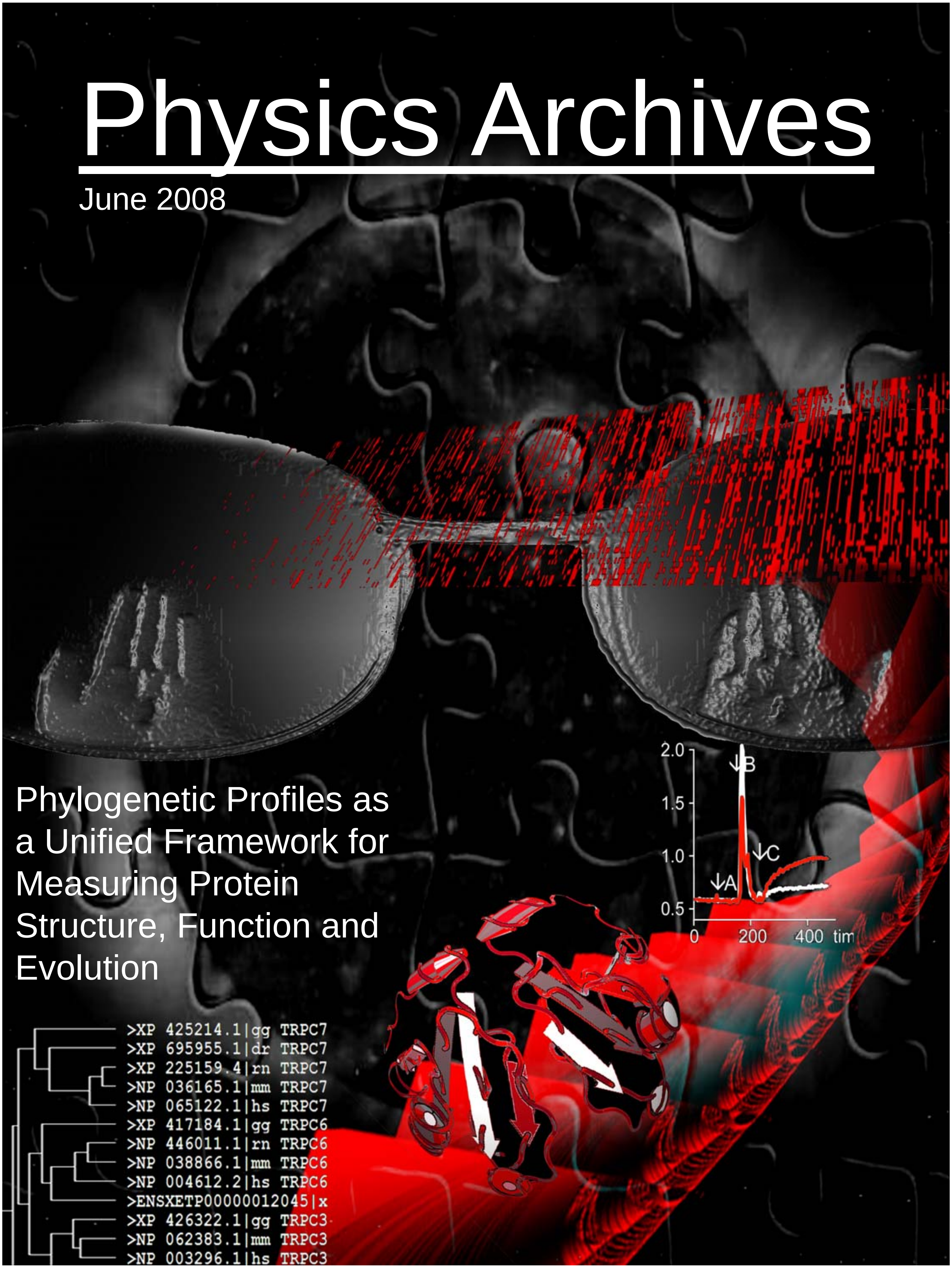

## Phylogenetic Profiles as a Unified Framework for Measuring Protein Structure, Function and Evolution

# Phylogenetic Profiles as a Unified Framework for Measuring Protein Structure, Function and Evolution

**Kyung Dae Ko[1†], Yoojin Hong[2†], Gue Su Chang[1†], Gaurav Bhardwaj[1], Damian B. van Rossum[1,3*], and Randen L. Patterson[1,3*]**

[1]Department of Biology, The Pennsylvania State University
[2]Department of Computer Science and Engineering, The Pennsylvania State University
[3]Center for Computational Proteomics, The Pennsylvania State University

[*] To whom correspondence may be addressed
[†] These authors contributed equally to this work

**The sequence of amino acids in a protein is believed to determine its native state structure, which in turn is related to the functionality of the protein. In addition, information pertaining to evolutionary relationships is contained in homologous sequences. One powerful method for inferring these sequence attributes is through comparison of a query sequence with reference sequences that contain significant homology and whose structure, function, and/or evolutionary relationships are already known. In spite of decades of concerted work, there is no simple framework for deducing structure, function, and evolutionary (SF&E) relationships directly from sequence information alone, especially when the pair-wise identity is less than a threshold figure ~25% [1,2]. However, recent research has shown that sequence identity as low as 8% is sufficient to yield common structure/function relationships and sequence identities as large as 88% may yet result in distinct structure and function [3,4]. Starting with a basic premise that protein sequence encodes information about SF&E, one might ask how one could tease out these measures in an unbiased manner. Here we present a unified framework for inferring SF&E from sequence information using a knowledge-based approach which generates phylogenetic profiles in an unbiased manner. We illustrate the power of phylogenetic profiles generated using the Gestalt Domain Detection Algorithm Basic Local Alignment Tool (GDDA-BLAST) to derive structural domains, functional annotation, and evolutionary relationships for a host of ion-channels and human proteins of unknown function. These data are in excellent accord with published data and new experiments. Our results suggest that there is a wealth of previously unexplored information in protein sequence, which when encoded using phylogenetic profiles can be harnessed to understand the fundamental properties of proteins.**

## Introduction

The protein problem has remained unsolved despite decades of heroic efforts[1,5]. In principle, one expects that protein sequence ought to determine structure, function, and evolutionary (SF&E) characteristics; while in reality, there still is no reliable method for predicting the native state structure of a protein or its function given only its primary amino acid sequence either experimentally or theoretically[1,5]. In addition, evolutionary measurements are stymied when homologous proteins are highly divergent from one another[6]. It is also well known that the number of putative protein sequences of any given length is enormous yet only a very few of these can be classified as proteins which fold rapidly and reproducibly and have useful function[5,7]. Despite this fact, there seems to be an astonishing simplicity to the protein problem because the number of distinct native state folds is extremely limited[7].

In general, the protein problem occurs due to the inability of current algorithms to identify homology between highly divergent protein sequences with statistical confidence. Moreover, sequence alignment can be unreliable for matching two sequences when the pair-wise identity is less than a threshold figure of about 25%, and alignments with lower identity (i.e. in the "twilight zone") are usually treated as random events[1]. However, a small number of conserved residues (~8% identity) can coordinate the 3D fold and/or function of proteins, with large portions of these proteins comprising heteromorphic pairs (i.e. protein sequences that form different folds depending on their sequence environment)[3,4,8]. This is likely because the key amino acids responsible for coordinating the fold ("signal") are preserved in evolutionarily related sequences, while less evolutionarily taxing amino acid substitutions ("noise") result in dilution of the identity signal below the threshold required by search algorithms to detect homology between sequences.

The fundamental question facing computational biologists is how can data spaces in the protein world (i.e. the structural, functional, and evolutionary spaces that proteins occupy) be accurately and quantitatively measured from primary amino acid sequences? Two philosophies can be used to approach this question. The first philosophy is to create tools which can accurately resolve a single aspect of SF&E as independent measures, followed by developing methods for relating these measures. This is the philosophy which has been the most pursued and has fostered the creation of excellent structural tools (Rosetta, MODELLER, Mustang, etc)[9–11], evolutionary tools (MSA programs [K-align, MUSCLE, CLUSTAL etc [12–14]] coupled to phylogenetic clustering algorithms [Neighbor-Joining, Maximum Likelyhood, Maximum Parsimony, etc [15–17]]), and functional tools (Pfam, SMART, CDD, Interproscan, etc)[18–21]. While all of these algorithms are quite amazing in their own right, there are no clear methods for quantitatively comparing the information from one measurement to another, and all of these methods are insensitive when homology cannot be defined due to lack of sequence identity.

A second philosophy is that only by measuring all aspects of SF&E, simultaneously, can homology truly be measured. Were this possible, (i) functional and evolutionary measurements could quantitatively inform structural modeling to derive accurate atomic resolution protein structures; (ii) structural and functional measurements could inform evolutionary histories to derive accurate evolutionary rates, deep-branch relationships, and homologous regions within each protein; and (iii) structural and evolutionary measures would inform as to the location of functionalities/regulatory-elements contained within any protein. Armed with this information, the speed at which diseases could be understood and pharmacophores/therapies developed to combat them would be significantly increased.

**Phylogenetic Profiles**

A phylogenetic profile of a protein is a vector, where each entry quantifies the existence of the protein in a different genome [22,23]. This approach has been shown to be applicable to whole molecule (Single Profile Method), to an isolated domain (Multiple Profile Method), and to individual amino acids. Further, phylogenetic profiles have previously been used to assign protein function as well as protein interaction partners[24–26]. In 2006 Shankar Subramaniam's group demonstrated the latter with a novel technique called Coevolutionary-Matrix[22,23]. They quantified co-evolution between regions of two proteins, exploiting residue-level conservation to identify co-evolving regions that might correspond to domains. Using *Escherichia coli* proteins, they determined that co-evolving regions of proteins detected by this method allows for the generation of hypotheses about their specific functions/protein-protein interactions, many of which are supported by existing biochemical studies. We demonstrate here that the use of phylogenetic profiles can be extended to measure SF&E within a unified and quantitative framework.

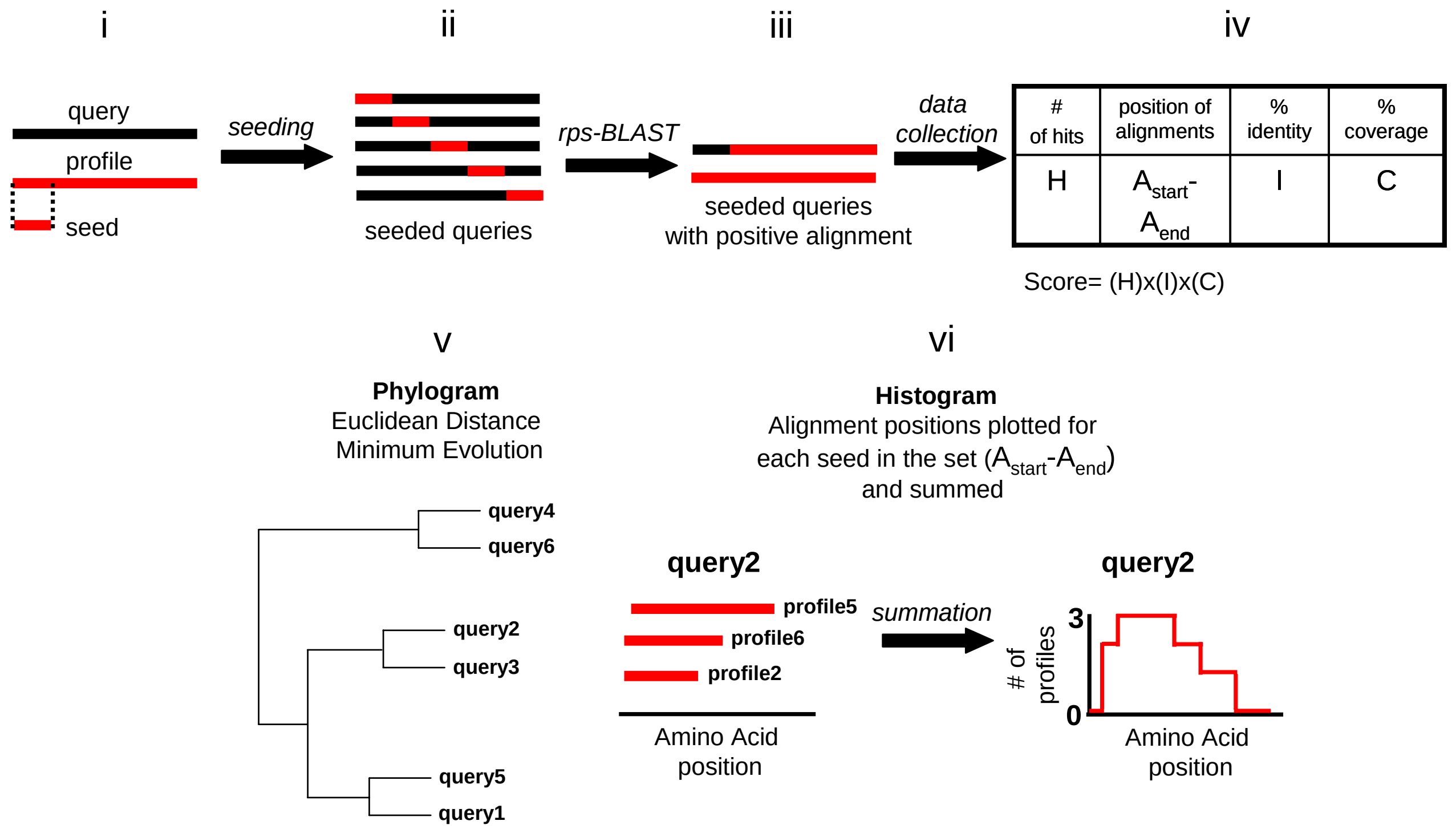

**Figure 1: GDDA-BLAST Concept.** The algorithm begins (i,ii) with a modification of the query amino acid sequence (black) at each amino acid position though the insertion of a "seed" (red) sequence from the profile. Each modified query sequence is then aligned by reverse position specific-BLAST (rps-BLAST)[20], against the profile sequence (iii). This process is repeated for all of the modified query sequences. We then filter the results of these alignments using simple thresholds such as % identity and % coverage (i.e. alignment length as a function of profile size) (iv). A composite score is then calculated as a product of (H)= hits, (I)= average % identity, and (C)= % coverage (iv). The composite score for all profiles tested are used to calculate the Pearson's Correlation Coefficient or Euclidian Distance between all pairs of query sequences (v). These distances can be visualized as a phylograms using standard algorithms (neighbor-joining, minimum evolution, etc) (vi). The alignments which are above threshold in (iv) can also be visualized as a histogram (see methods).

## GDDA-BLAST Encodes Phylogenetic Profiles

GDDA-BLAST matrices are a variation of phylogenetic profiles, except in our case, a protein is a vector where each entry quantifies the existence of alignments with a sequence domain profile. To generate phylogenetic profiles using GDDA-BLAST, one begins by compiling a set of profiles that the query sequence is compared to. These profiles can be obtained from multiple knowledge-base sources (e.g. Pfam, SMART, NCBI Conserved Domain Database (CDD))[18–20]. We employ reverse specific position BLAST (rps-BLAST) to compare query and profiles sequences[20]. Rps-BLAST is a two step algorithm which first finds an initiation location with highly similar amino acids in both the query and profile sequence, followed by extension of the alignment[20]. Often rps-BLAST is unable to discover alignments when the query sequence is divergent from the profile sequence. To overcome this limitation we have introduced multiple innovations in GDDA-BLAST.

We utilize a single domain profile database for pairwise comparisons. Then we record and quantify all alignments between an unmodified (control), and modified query sequence. The latter is composed of two types of alignments: "seeded" and non-seeded alignments. We modify the query sequence with a "seed" from the profile, creating a consistent initiation site. This consistent initiation site allows rps-BLAST to extend an alignment even between highly divergent sequences[27–32]. This resampling strategy is designed to amplify and encode the alignments possible for any given query sequence. Instead of a sliding window, we utilize a sliding "seed".

"Seeds" can be generated from a profile by taking a fraction (e.g. 3-50%) of the profile sequence (e.g. starting from the N-terminus, or from the middle, or ending at the C-terminus). These "seeds" are inserted at each location of the query sequence one at a time. Thus, a query sequence of N amino acids yields N distinct test sequences for each "seed" (Fig 1-ii). Each of these modified [test] sequences is aligned by rps-BLAST[20] against the parent profile (Fig 1-iii). This allows rps-BLAST to extend the alignment even between highly divergent sequences. Next, we filter the results using simple thresholds such as % identity and % coverage (i.e. alignment length as a function of profile length) (Fig 1-iv). This procedure can be readily adapted to make an unbiased comparison between a series of query sequences by subjecting them to the same screening analysis with the same set of profile sequences as "seeds" (Fig 1-v).

The phylogenetic profile is generated by representing each query sequence as a vector of non-negative numbers in M dimensions (M= # of "seeds" tested). These profiles can then be used to create a tree of relationships based on standard statistical techniques such as Euclidian distance or Pearson's correlation between each query sequence (Figure 2F-I)[30,31]. The success of the method, as shown below, lies in the choice of profiles geared towards the specific question one wishes to answer.

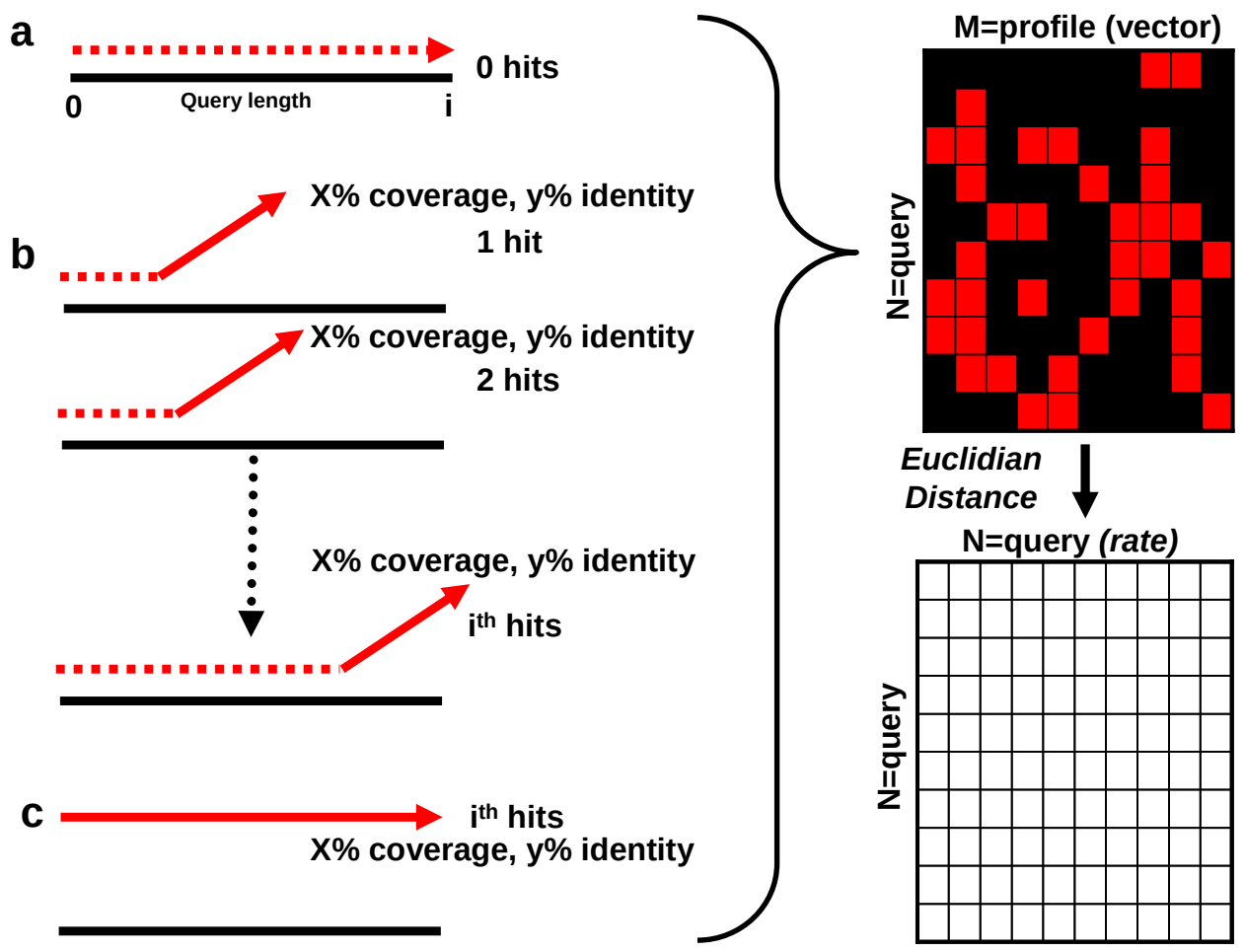

**Figure 2: GDDA-BLAST phylogenetic profiles encode rate information.** When a series of "seeded" queries is aligned against a given profile in rps-BLAST, one of three results can be returned; (a) no result, (b) a number of hits [alignments] above identity and coverage thresholds, or (c) hits at every position above thresholds. Thus, the number of hits is equivalent to force (e.g. 100 hits is more forceful than 10 hits), with the coverage and identity providing the spatial coordinates for the force (e.g. coverage = x-axis, identity = y-axis). (d) These vectors can be encoded within an N(query) X M(profile) matrix (see Methods). This matrix can then be compared all-against-all using Euclidian Distances to create a NXN matrix (see Methods). The distances in this NXN matrix are proportional to the rate of divergence between the query proteins.

## Evolutionary rates derived from GDDA-BLAST phylogenetic profiles

In order to determine evolutionary relationships between homologous proteins requires measurements of evolutionary rates. We propose that rate information can be measured from a phylogenetic profile. As shown in Figure 2a-c, GDDA-BLAST phylogenetic profiles are encoded as vectors. Each "seeded" query can return either no alignment, or an alignment that ranges over %identity and %coverage; thus we encode the N X M matrix (Figure 2d) with these vectors. A Euclidian distance can then be generated from this N X M vector matrix (see Methods) with the simple assumption that the distance between each N [query] in the matrix is proportional to the rate of evolutionary divergence.

Indeed, Figure 3 shows the results of our characterization of 20 water-channel (aquaporin) proteins with 23,605 profiles from the NCBI CDD database[20]. We find that there are four distinct families with rates that accord with previous studies employing multiple sequence alignment[33]. From random considerations, the probability of organizing these twenty sequences correctly into 4 families is $9X10^{-13}$. These results demonstrate that phylogenetic profiles derived by GDDA-BLAST can contain evolutionary rate information, which is independent of multiple sequence alignment based methods. We believe that rigorous analyses on benchmark training sets will enable us to make more refined and statistically robust measurements among distantly related and/or rapidly evolving proteins.

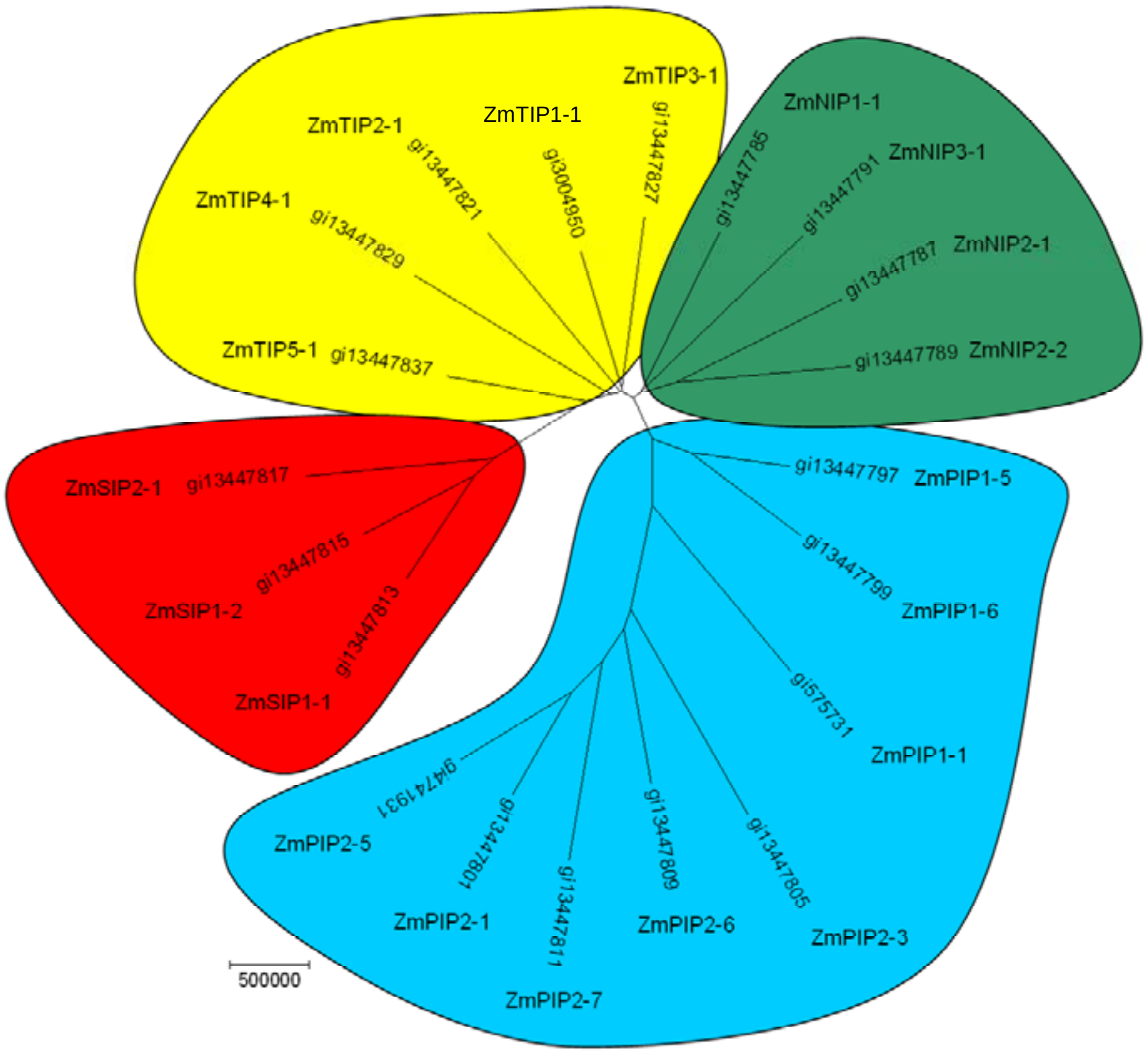

**Figure 3: Water Channel (Aquaporin) Phylogeny.** Twenty *Zea Maize* aquaporin channels (plasma membrane intrinsic proteins (PIPs), tonoplast intrinsic proteins (TIPs), Nod26-like intrinsic proteins (NIPs), and small and basic intrinsic proteins (SIPs)) were screened with GDDA-BLAST (see methods). The Euclidian distance is generated from the composite scores and plotted in an unrooted tree using the MEGA3 minimum evolution algorithm[17]. Scale bar reflects the Euclidian distance between sequences and color coding reflects the distinct and known classes of aquaporins. Our results are in excellent accord with the findings of Chaumont et al[33].

## Structure homology obtained using GDDA-BLAST phylogenetic profiles

It has been proposed that the number of distinct native state folds is extremely limited[5,7]. This suggests that with accurate measurements of homology, inferences of structure from primary amino acid sequences are possible. To evaluate the ability of GDDA-BLAST phylogenetic profiles to measure homologous relationships, we present here a performance test using the SABmark "twilight zone" set of proteins curated from the SCOP structural database (see Methods).

Figure 4 shows ROC curves produced by GDDA-BLAST (employing Pearson's correlation coefficient similarity metric), SAM-T2K, and PSI-BLAST to detect homologous sequences from 534 sequences representing 61 groups (average sequence length: 135.27 ± 83.39 s.d.) which were randomly selected from the SABmark twilight zone set [34]. We observe that GDDA-BLAST outperforms both algorithms at statistically significant thresholds (0.01 and 0.05). Our data also demonstrates that addition of knowledge-base profiles increases GDDA-BLAST performance (compare 8096 vs. 11101 profiles). Importantly, GDDA-BLAST results were derived with only a small subset of knowledge base profiles and at default settings which can be optimized.

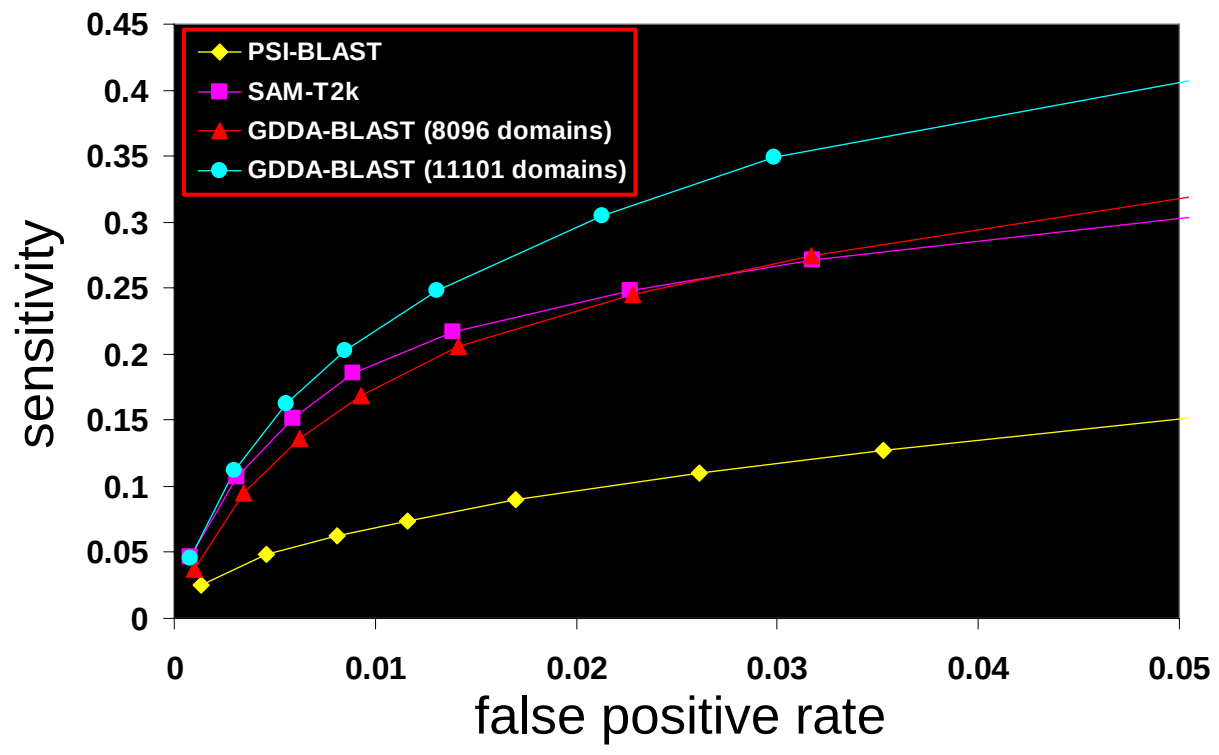

**Figure 4: ROC Curves on the "Twilight Zone" SABMARK Benchmark Dataset.** For GDDA-BLAST, all protein pairs are encoded as a vector of composite scores (%coverage * %identity * # of hits) for 8096 CDD domain profiles (all profiles in the CDD< 144 amino acids) and for 11101 CDD profiles (added 3005 profiles between 144 and 189 aa). As above, each of the sequences is compared in an all-against-all fashion. To compare encoded sequences, Pearson's correlation coefficient is computed to measure the similarity between two sequences. All values for the Pearson's correlation are accepted into the comparative analysis. ROC performance analysis were performed by top-K analysis (Bradley, 1995) with largest correlation coefficient or smallest e-value. This sequence is considered homologous sequences to the query.

## Structural boundaries of ion-channels

A key structural feature of channels is their transmembrane helices which comprise their pore-forming domain. We demonstrate that GDDA-BLAST has the capacity to model these domains in multiple classes of ion-channels. Results of GDDA-BLAST using 98 transmembrane containing profiles from the CDD database accurately isolate the channel domain in the plant aquaporin PIP2-6, and other ion channel domains as well (Fig 5). All of our results are in accord with those obtained using the Hidden Markov Model, TMHMM (see methods). Indeed, our measurements are robust even when the transmembrane regions of the channel are separated by an intervening sequence, as is demonstrated by the cystic fibrosis transmembrane conductance regulator (CFTR) (Fig 5a, right). By comparison, rps-BLAST alone poorly annotates channel boundaries (see Methods) demonstrating that GDDA-BLAST vastly increases the predictive power of general profile-based searches.

A recent study by Mio et al. obtained a cryo-EM structure of Transient Receptor Potential Channel 3 (TRPC3) and modeled the six transmembrane helices with the atomic structure of the potassium channels KcsA and Kv1.2 [35]. Interestingly, these authors also determined that TRPC3 contains a globular, and presumably hydrophobic, inner-core surrounded by signal sensing antenna derived from the cytosolic N and C-termini (Fig 6a). We wondered whether these channel constituents could be computationally modeled with GDDA-BLAST, by generating phylogenetic profiles

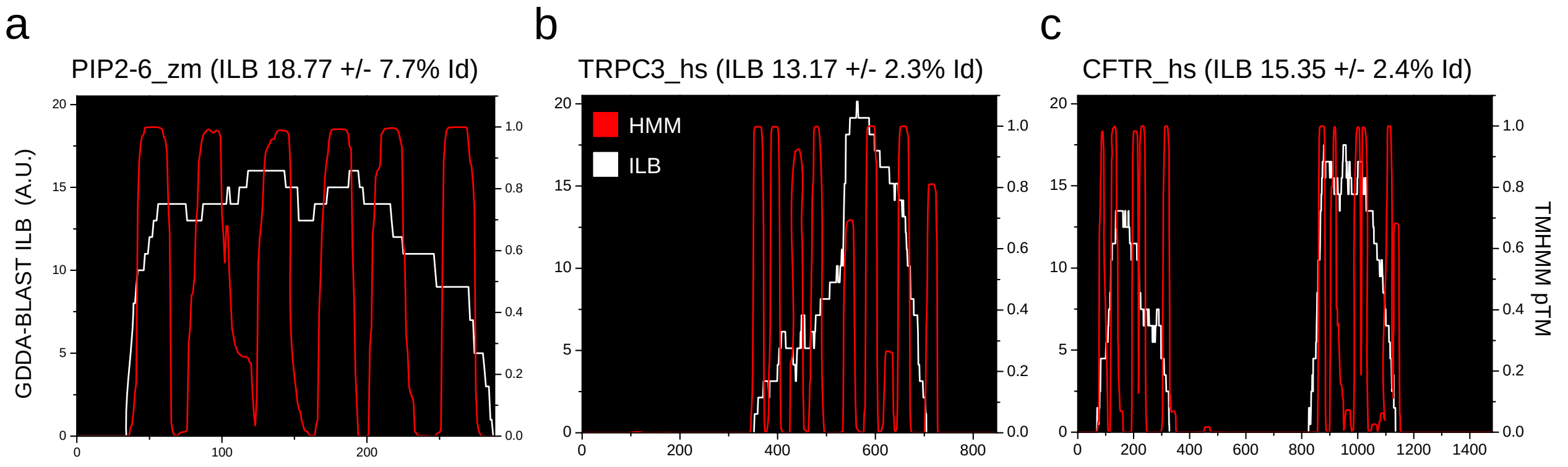

**Figure 5: Annotation of functional and structural domain boundaries.** (a-c) GDDA-BLAST results for the screen of ZmPIP2-6, TRPC3, and CFTR respectively with 98 integral lipid-binding profiles (see methods). The results from this screen (left axis, white) are plotted together with transmembrane probabilities derived from TMHMM analysis (right axis, red), showing excellent accord.

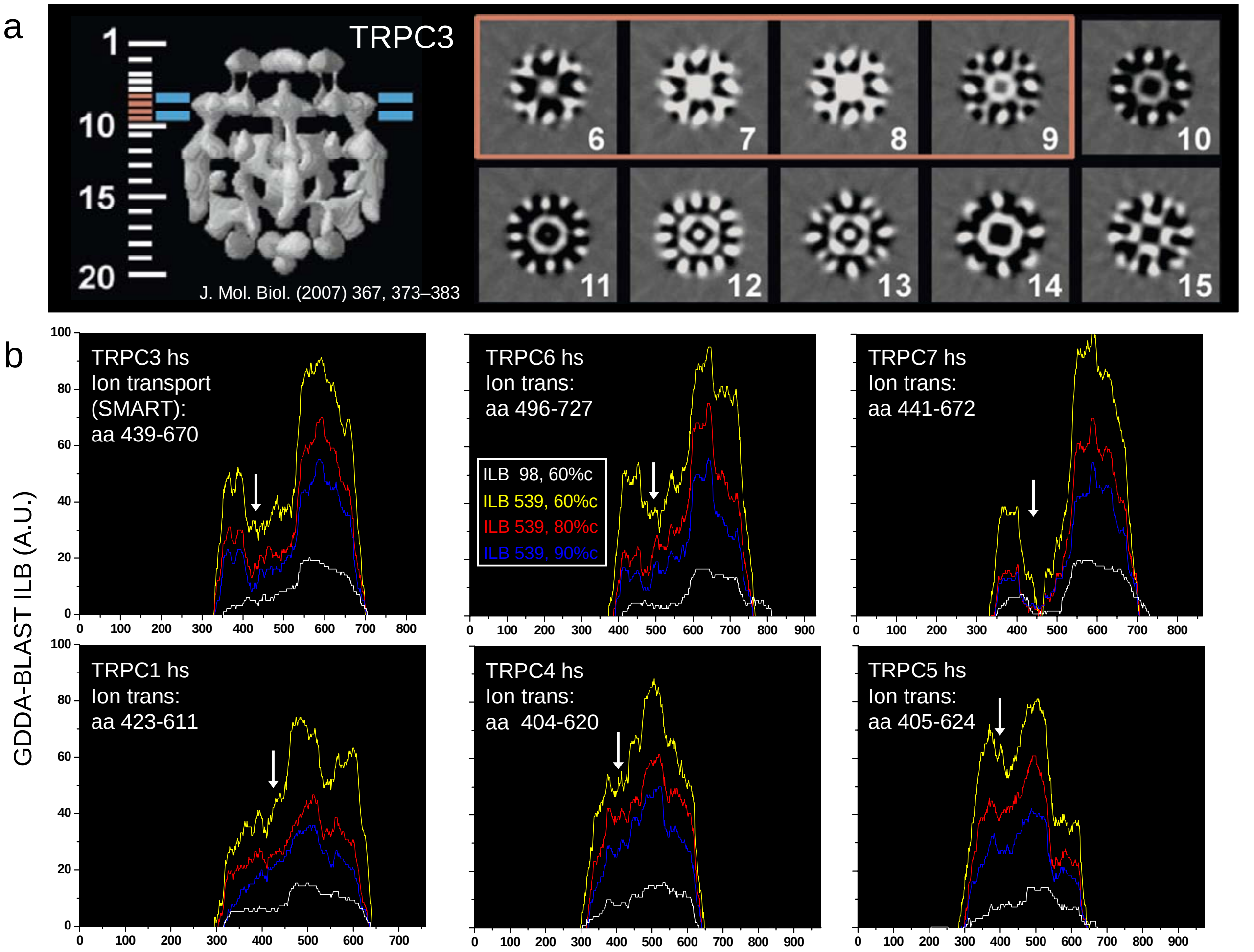

**Figure 6: GDDA-BLAST models of the ion transport domain in TRPC channels.**
(a) 3D reconstruction of TRPC3 channel derived by Mio et al. (J. Mol. Biol. (2007) 367, 373–383). Blue lines depict the plasma-membrane. The scale on the left depicts the cryo-electron microscopic images of horizontal slices parallel to the plasma-membrane (images 6-9) progressing into the cytosol (images 10-15). The globular inner shell can be seen as a circular density in the center of the images. (b) GDDA-BLAST results for human TRPC channels using 98 curated integral lipid-binding (ILB) profiles and 576 profiles parsed with key words for (channel, transmembrane, integral membrane, and/or pump). The latter were also analyzed with different % coverage thresholds. Ion transport boundaries in TRPC channels predicted by SMART (default settings) are noted with the N-terminal boundary denoted by an arrow. GDDA-BLAST results predict that the globular inner shell domain is located to the left of the arrow.

from sequences that comprise the appropriate structural elements/biological functions of interest.

Initially, we queried human TRPC channels with a curated set of 98 transmembrane domain containing profiles to generate our GDDA-BLAST phylogenetic profiles (see Methods). The distribution of the alignments which are above threshold is plotted in Fig 6b, white. The results from this experiment accurately model the channel domain in human TRPC channels when compared with transmembrane predictions by the hidden Markov model TMHMM and the domain detection algorithm SMART [19,36]. We tested whether key-word searches of the NCBI CDD database (CDD) could be used to collect additional points of information to our phylogenetic profiles. We collected 536 profiles in CDD which have the following key words (channel, transmembrane, integral membrane, pump) and performed our analysis again (Fig 6b, yellow). We observe that alignments against these profiles also model the channel domain boundaries. In addition, a pronounced peak is evident in TRPC3/6/7 that significantly differs in TRPC1/4/5. This signal likely represents the hydrophobic globular inner-core domain in TRPC3 identified by Mio et al., and suggests that the channel domains in TRPC1/4/5 are likely different structurally and/or functionally from TRPC3/6/7. To determine whether these signals are robust, we recalculated the data using % coverage thresholds ranging between 60% and 100% (Fig 6b, red and blue). Surprisingly, a 60% threshold does not significantly alter the domain boundaries, but does increase the signal in our results. Overall, the GDDA-BLAST model of TRPC ion-channel domains is in excellent accord with other computational models and experimental evidence.

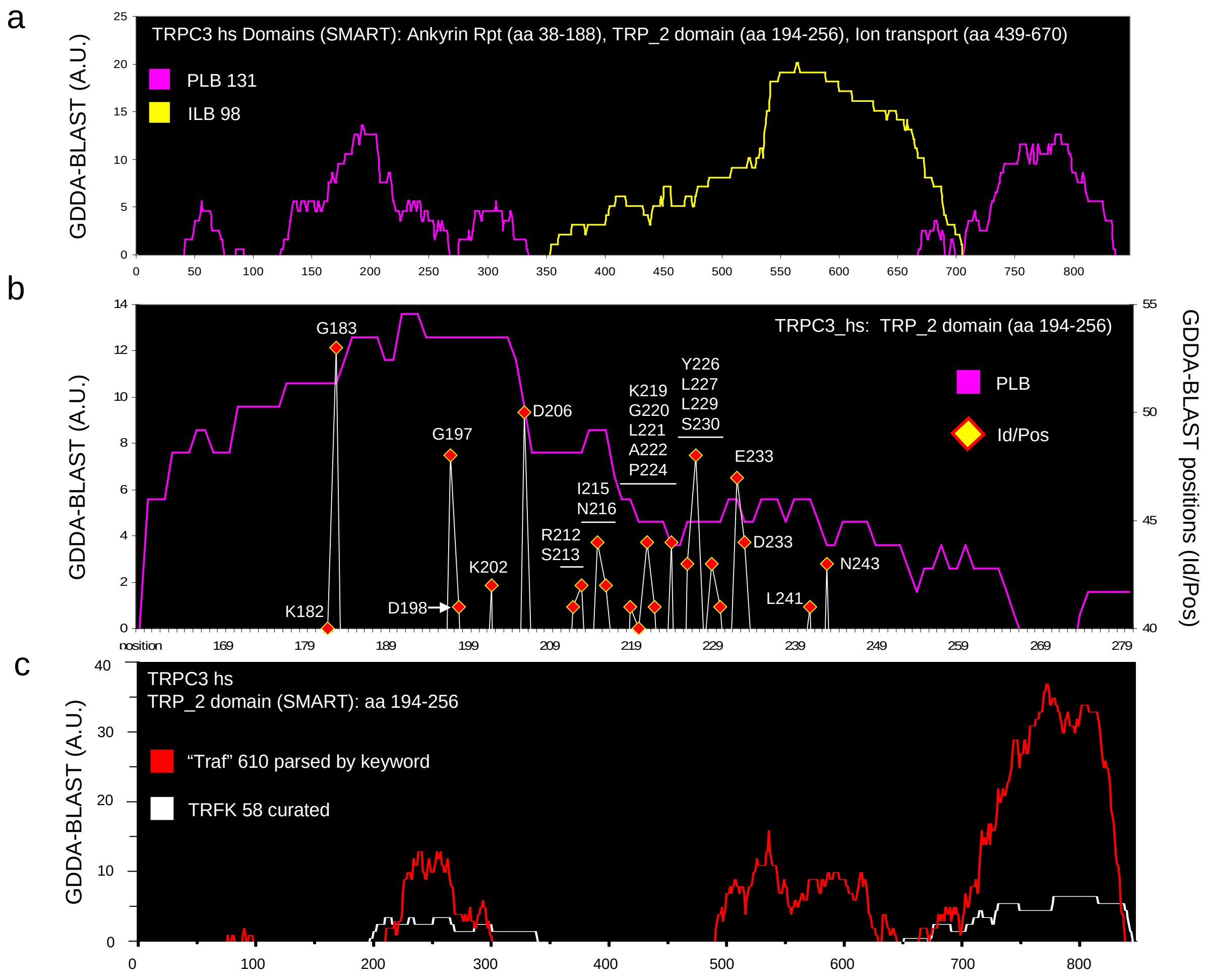

**Figure 7: GDDA-BLAST models of the lipid-binding domains in TRPC3.** (a) Domain boundaries in TRPC3 predicted by SMART (default settings) are noted. GDDA-BLAST results for human TRPC3 channel using 131 peripheral lipid-binding (PLB), and 98 integral lipid-binding (ILB) profiles. (b) (Right y-axis): GDDA-BLAST results in the region of the TRPC3 TRP_2 domain with the 131 peripheral lipid-binding profiles overlaid with (Left y-axis): the quantification of amino acid positions which are identical or similar in alignments with peripheral lipid-binding profiles (see Methods). (c) GDDA-BLAST results for human TRPC3 channel using 58 curated trafficking profiles (TRFK), and 610 profiles which were parsed by the keyword "traf". These trafficking profiles both give a signal in the TRP_2 domain and in the C-terminus, regions of which are also positive for peripheral lipid-binding profiles.

## TRP_2 domain Modeling: *Peripheral lipid-binding and Trafficking*

Given the success of the aforementioned channel modeling, we wondered whether GDDA-BLAST could be used to model other functional domains. TRPC3 has two known lipid-binding domains which are not predicted by conventional domain detection algorithms. We generated GDDA-BLAST phylogenetic profiles using 131 peripheral lipid-binding profiles from CDD. We observe multiple peaks in the histogram of TRPC3 (Fig 7a, pink). Importantly, we obtain peaks in both regions of TRPC3 that have been experimentally validated to bind lipid [27,37]. We also observe an additional peak in the domain of unknown function, TRP_2. We investigated the pair-wise alignments obtained from GDDA-BLAST and determine the frequency of specific amino acids (within the TRP_2 domain) aligning with specific functional profiles (Fig 7b, Methods). This type of strategy is analogous to a sequence logo[38]. We observe that most of the conserved residues from this analysis occur between G197 and D233. We then tested the prediction that TRP_2 binds peripheral lipids in the physiologically relevant liposomal assay constructed to mimic plasma-membrane lipid composition[39]. In these *in vitro* assays we observe that the TRP_2 domain of TRPC3 binds to lipids and that mutations in conserved residues alter this function (*manuscript submitted Jounal of Biological Chemistry*).

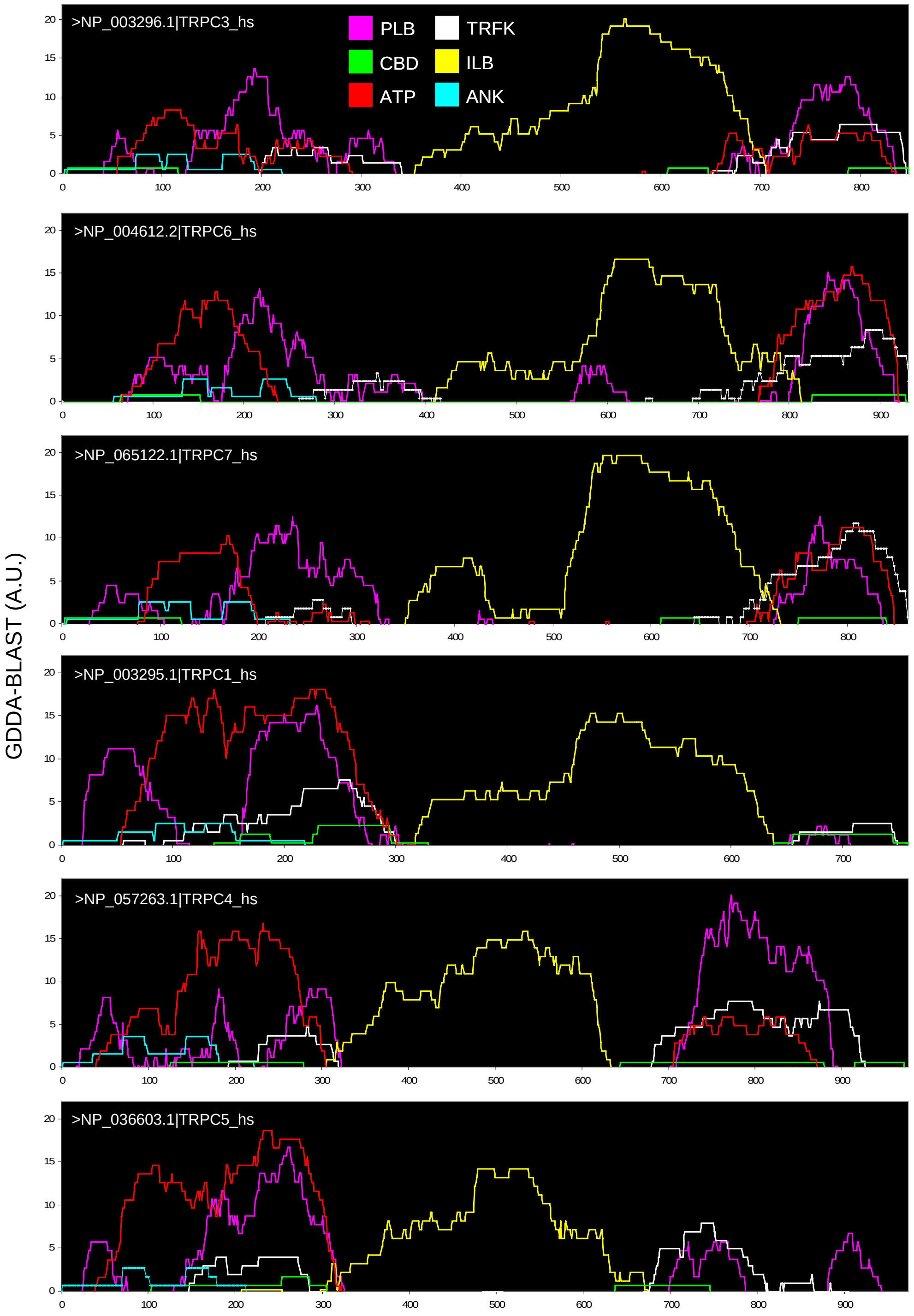

**Figure 8: GDDA-BLAST models of human TRPC channels.** GDDA-BLAST results for human TRPC channels using 131 peripheral lipid-binding (PLB), 98 Integral lipid-binding (ILB), 58 Trafficking (TRFK, n=58), 10 Calmodulin-binding (CBD), 4 Ankyrin Repeat (ANK), and 574 ATP (ATP) profiles.

Mutations to the TRP_2 domain which alter lipid-binding also alter channel function and/or trafficking. We generated phylogenetic profiles for TRPC3 using a curated set of 58 protein trafficking profiles from CDD (Fig 7c, white). We observe a signal for trafficking profiles in the TRP_2 domain and in the C-terminus, both regions of which are also positive for peripheral lipid-binding profiles (see Fig 7a). We then used the key-word "traf" (traffic, trafficking, and trafficked) to collect 610 additional profiles from CDD and tested them with GDDA-BLAST (Fig 7c, red). We obtain similar signals from these phylogenetic profiles, although more robust, to those obtained with the hand-curated trafficking set. Our experimental evidence confirmed our model since altering TRPC3 lipid-binding also alters TRPC3 fusion with the plasma-membrane in response to diacylglygerol (*manuscript submitted Jounal of Biological Chemistry*).

**Ankyrin-repeat modeling in TRP channels**

It is well established that ankyrin repeats can perform a number of functions (e.g. ATP-binding, lipid-binding, calmodulin-binding)[40–42]. However, to our knowledge there are no current domain-detection algorithms which can resolve their multi-functional nature. Therefore, we generated multiple phylogenetic profiles for the human TRPC channels using profiles obtained from CDD with the following key words (ankyrin [4 profiles], ATP [574 profiles], calmodulin-binding [10 profiles]) (Fig 8a). We observe signals for all of these profiles within the ankyrin repeats of all human TRPC channels at varying levels of intensity. Importantly, we observe signals for calmodulin-binding in the C-terminus of all TRPC channels, which accords with findings by Kwan et al [37]. These results predict that TRPC channels are regulated by calmodulin through multiple binding interactions, a result which is in accord with the literature and has been observed in other ion-channels[43,44].

To provide confidence for our predictions of ATP-binding within the ankyrin repeats of TRPC channels, we turn to the vanilloid TRP (TRPV) family. Lishko et al. recently crystallized the ankyrin repeats of TRPV1 and TRPV2 and found their structures to be highly similar [40]. Interestingly, these authors also determined that both ankyrin repeats bound to calmodulin, while only TRPV1 was capable of binding ATP in their assays. Indeed, when we obtain phylogenetic profiles for TRPV1 and TRPV2 with GDDA-BLAST exactly as the human TRPC channels, we observe calmodulin signals in the ankyrin repeats of both TRPV1 (Fig 9a) and TRPV2 (data not shown). Conversely, TRPV1 has a robust ATP signal within its ankyrin repeats while TRPV2 is significantly smaller (only 18% of TRPV1, area under the curve) (Fig 9b). When the conserved residues from the ATP profile alignments are plotted (see Methods), we observe that the top scoring residue in TRPV1 is E211, which coordinates the N6 amine binding of ATP in the active pocket (Fig 9c). All of these results are in excellent accord with the experimental evidence of Lishko et al [40]. Of note, when we quantify the ATP signal in multiple ankyrin repeats, it varies over a wide range suggesting that ATP-binding in ankyrin repeats is quite variable, even within closely related proteins, such as TRPC channels.

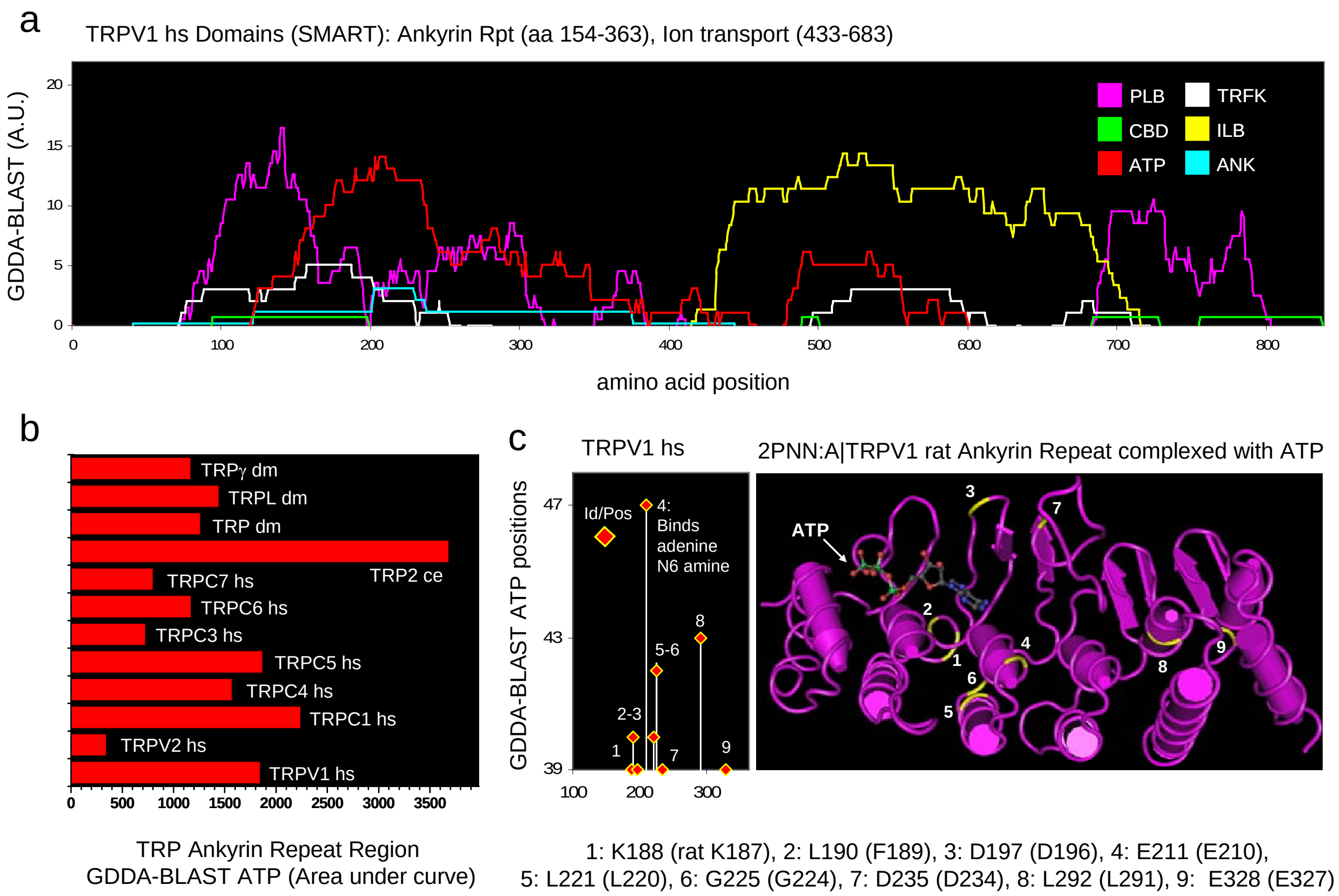

**Figure 9: GDDA-BLAST model of the ATP-binding Ankyrin Repeat in TRPV1.** (a) GDDA-BLAST results for human TRPV1 channel using 131 peripheral lipid-binding (PLB), 98 Integral lipid-binding (ILB), 58 Trafficking (TRFK, n=58), 10 Calmodulin-binding (CBD), 4 Ankyrin Repeat (ANK), and 574 ATP profiles. (b) GDDA-BLAST results for the screen of 574 ATP profiles in the ankyrin repeat domain of various TRP channels was integrated to quantify the area under the curve and plotted in a bar graph. (c) Left: Quantification of amino acid positions in human TRPV1 ankyrin which are identical or similar in alignments with ATP profiles (see Methods). Right: Crystal structure of the rat TRPV1 ankyrin repeat complexed with ATP (PDB: 2PNN). Residues depicted in yellow are homologous to those derived in human TRPV1.

## Uncovering novel lipid-binding domains *in vitro*

Structure and function can be independent (e.g. PH-superfolds: PH-domain, EVH domain, Ran-binding domain= function independent) and functionally related domains can have related motifs (e.g there are only so many chemical combinations of amino acids to coordinate an enzymatic and/or binding pocket)[7,45]. Our results with transmembrane and trafficking domains display a phenomenon whereby alignments for domains of similar function and diverse structures coalesce. We suggest this phenomenon occurs due to the encoding of this information within our phylogenetic profiles. To address this supposition, we performed an additional test on human proteins of unknown function. These proteins were chosen specifically for their lack of functional annotation by any current algorithm tested (see Methods).

The peripheral lipid-binding profiles used in the above assays have binding specificity for a wide range of lipids (e.g. phospholipids, terpenes, fatty-acids, etc.). We observe multiple peaks in the histograms generated from these alignments (Fig 10a). Next, we cloned representative regions from each of these proteins and prepared bacterially purified protein. These purified proteins were subjected to liposomal assays containing lipids which mimic the plasma-membrane of animal cells (see Methods). Strikingly, each of the fragments containing GDDA-BLAST signals was positive for lipid-binding, whereas our negative controls were not (Fig 10b). Although the physiological relevance of these lipid-binding domains remains to be determined, these results clearly demonstrate that phylogenetic profiles generated using ontological relationships are effective for identifying putative functions within protein domains.

## Discussion

In summary, we present a new tool for using phylogenetic profiles to simultaneously infer SF&E relationships of proteins. GDDA-BLAST is a reliable technique for unbiased analysis of protein sequences with a carefully chosen set of profiles. Our computational study, validated by experiment, suggests that, even in its nascent stage, GDDA-BLAST is a promising technique for decoding the proteomes of organisms. Further, the general use of "seeding" as an re-sampling technique may be useful in other contexts of pattern recognition such as artificial intelligence[46], weather prediction[47], economic indicators[48], and seismographic movements[49].

GDDA-BLAST can accurately model structure/function relationships in TRP channels. This is supported by our findings that GDDA-BLAST predicts: (i) the ion-channel domains of TRP channels, (ii) lipid-binding and trafficking function within the previously uncharacterized TRP_2 domain, and (iii) the multi-functional (lipid-, calmodulin-, and ATP-binding) nature of ankyrin repeats within TRP channels. Our experimental evidence demonstrates that TRPC3 TRP_2 is a lipid/trafficking domain that contributes to DAG-sensitive vesicle fusion (*manuscript submitted Journal of Biological Chemistry*). GDDA-BLAST models of TRPC channels also recapitulate experimental evidence from multiple laboratories. For example, the homologous C-terminal domain of TRPC6 was recently reported to bind both $PIP_3$ and calmodulin in various ion channels yet is undetectable by conventional methods [37].

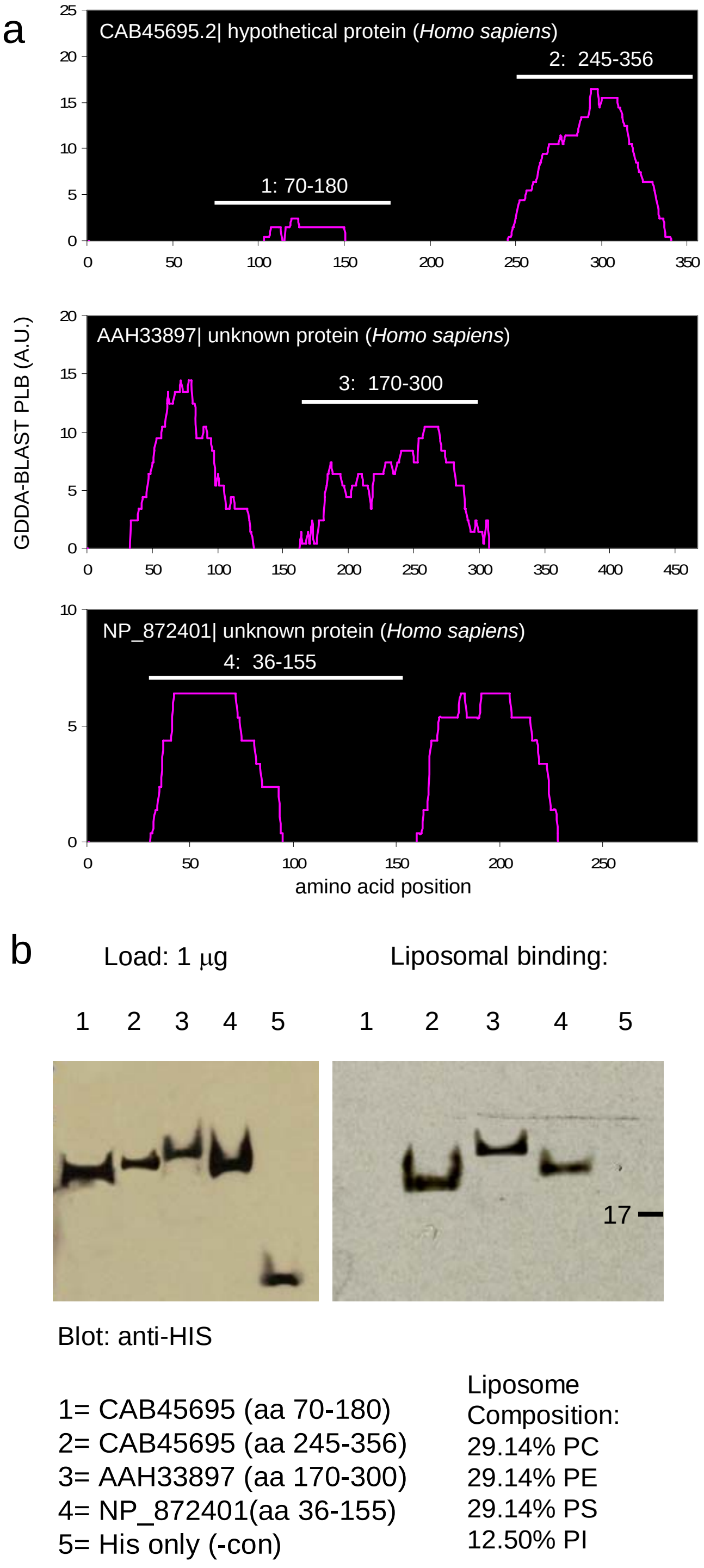

**Figure 10: Functional Information via GDDA-BLAST analysis**
(a) GDDA-BLAST results for three human proteins of unknown function (AAH33897, NP_872401, and CAB45695.2) using 131 peripheral lipid-binding (PLB) profiles. The white bars depict regions that we cloned for liposomal experiments in (b). (b) Western analysis of purified CAB45695, AAH33897, NP_872401, fragments cloned into His vector (1 mg load). These fragments were tested for binding to liposomes containing phosphatidylcholine (PC), phosphatidylethanolamine (PE), phosphatidyl serine, and phosphatidylinositol (PI). All fragments bound to liposomes except fragment 1 (CAB45695: aa 70-180) and the HIS-tag in perfect accord with the predictions of GDDA-BLAST.

GDDA-BLAST readily predicted this domain and its functions. GDDA-BLAST also accurately models the ATP-binding activity contained in the ankyrin repeats of the structurally resolved TRPV channels[40]. We also observe a segmented signal in TRPC3/6/7 when tested by GDDA-BLAST with transmembrane domain profiles, which likely represents the globular inner-core domain observed in the cryo-EM structure obtained by Mio et al. GDDA-BLAST also predicts that likely all plasma-membrane resident ion-channels contain peripheral-lipid binding and trafficking domains, based as we also observe multiple lipid-binding domains in all channels tested (e.g. aquaporins, and $Na^+$, $K^+$, $Cl^-$, $Ca^{2+}$ channels) (data not shown, available upon request). All of these channels have been demonstrated, empirically, to interact with lipids[50].

We propose that GDDA-BLAST measurements can be treated as "fingerprints" of SF&E information. Through careful choice of knowledge-base profiles related for either structural or functional qualities, GDDA-BLAST provides results which can be used to infer evolutionary rate information, create functional models and identify structural boundaries for protein sequences, even if no prior information exists. Perhaps most important, GDDA-BLAST has the capacity to inform laboratory experiments of key amino acids important to protein function, thus speeding the discovery process. Our studies here demonstrate one way of using phylogenetic profiles to quantitatively probe knowledge-bases to obtain SF&E information within the same unified framework. Future work aimed at determining the data points collected by GDDA-BLAST which are informative for SF&E annotation, and which ones are sufficiently noisy such that they are detrimental to the total information content will enable us to understand and harness the underlying mechanisms of our algorithm thus optimizing and refining our approach.

**Acknowledgements**: This work was supported by the Searle Young Investigators Award, start-up monies from PSU (RLP), NCSA grant TG-MCB070027N (RLP, DVR), BIO060003P from the Pittsburgh Supercomputing Center (RLP) and the Thomas O. Magnetti fund (RLP/DVR). This project was also funded by a Fellowship from the Eberly College of Sciences and the Huck Institutes of the Life Sciences (DVR) and a grant with the Pennsylvania Department of Health using Tobacco Settlement Funds (DVR),. The Department of Health specifically disclaims responsibility for any analyses, interpretations or conclusions. Thanks go to Jason Holmes at the PSU CAC center for technical support. Special thanks go to our friend and colleague Jayanth Banavar, for the generous gift of his time, unerring intellectual critique and encouragement, and for invaluable help with the construction of this manuscript. We also thank Jim White, Barbara VanRossum, Solomon H. Snyder, Kirill Kiselyov, Robert E. Rothe, Curtis Jackson, Lloyd Banks, Y. Buck, J.R. Baloo, Lisa Renyolds, Wojtek Makalowski, R.I. Peace, T. Shakur, and B. Smalls for creative dialogue.

**Contact Information**

Randen L. Patterson
230 Life Science Building
Department of Biology
The Pennsylvania State University
University Park, PA 16802, USA
Tel: 001-814-865-1668
Fax: 001-814-863-1357
e-mail: rlp25@psu.edu

Damian B. van Rossum
518 Wartik Labs
Department of Biology
The Pennsylvania State University
University Park, PA 16802, USA
Tel: 001-814-863-1007
e-mail: dbv10@psu.edu

## Materials and Methods

**Ion channel sequences used in this study-**
>gi|4507685|ref|NP_003295.1| TRPC1 [Homo sapiens],
>gi|4507687|ref|NP_003296.1| TRPC3 [Homo sapiens],
>gi|7706747|ref|NP_057263.1| TRPC4 [Homo sapiens],
>gi|6912736|ref|NP_036603.1| TRPC 5 [Homo sapiens],
>gi|5730102|ref|NP_004612.2| TRPC 6 [Homo sapiens],
>gi|9966865|ref|NP_065122.1| TRPC7 [Homo sapiens],
>gi|74315350|ref|NP_061197.4| TRPV1 [Homo sapiens],
>gi|20127551|ref|NP_057197.2| TRPV2 [Homo sapiens],
>gi|71989165|ref|NP_001022703.1| TRP-2 [Caenorhabditis elegans]
>gi|17136554|ref|NP_476768.1| TRP [Drosophila melanogaster],
>gi|24584649|ref|NP_609802.1| TRP-gamma [Drosophila melanogaster],
>gi|24652173|ref|NP_724822.1| TRPL [Drosophila melanogaster],
>gi|90421313|ref|NP_000483.3| CFTR [Homo sapiens].

**GDDA-BLAST Histograms-** In order to generate histograms which reflect structural boundaries and putative function, the profiles of related structure or function are curated. Each of the query sequences is compared to the chosen profiles using GDDA-BLAST. The distribution of the number of hits, above a predetermined threshold and summed over all the profiles, is determined at each position of the query sequence for all of these alignments. We chose to normalize the data by subtracting the mean number of hits per amino acid and the histograms in Figs 5-10 show just the positive scoring regions.

**GDDA-BLAST Phylograms-** Each aquaporin was screened using GDDA-BLAST with 23,605 profiles with two “seeds” each from the N-terminal and C-terminal regions whose length is 10% of the profile. For each profile scoring above threshold (60% coverage including the “seed”, 10% identity excluding the “seed”), a composite score is generated (% coverage * % identity * # of hits), creating an N (queries) by M (profile) matrix. The Euclidian distances between all pairs of query sequence composite scores is calculated creating an N (queries) by N (Euclidian distance) matrix. This matrix can then be analyzed using any phylogenetic method (e.g. Minimum Evolution, Neighbor Joining, UPGMA, etc).
Sequence accession numbers for these channels can be found in Figure 3.

**GDDA-BLAST Positional Analysis-** The graphs in Fig 7 and 9 were calculated by aligning the predicted domain boundaries in the query sequence against those profiles which were positive (>/= 60% coverage including the “seed” and >/= 10% Identity excluding the “seed”) in GDDA-BLAST. This was performed using EMBOSS with the settings: water (local), Blosum62, GOP: 5 and GEP: 1. Residues were scored value=2 for identities and value=1 for positive substitutions. These positions were tallied and the cumulative score was annotated versus the amino acid position. Fig 7b and 9c depict only the top scoring positions.

**Channel Domain analysis-** Thirteen transient receptor potential channels from the NCBI protein database (unless otherwise noted) (>c422101106.Contig1 (http://evodevo.bu.edu/stellabase/), NP_001022703.1, ENSCINP00000017401 (http://www.ensembl.org/), NP_476768.1, NP_609802.1, NP_724822.1, NP_035774.1, NP_003295.1, NP_003296.1, NP_004612.2, NP_036603.1, NP_057263.1, NP_065122.1) were analyzed using TMHMM to serve as our control. The same 13 channels were analyzed using rps-BLAST as well as GDDA-BLAST. Predictions of the N-terminal and C-terminal boundaries for transmembrane domains predicted were recorded. GDDA-BLAST results were in excellent accord with control TMHMM (110 +/-16% of control), while rps-BLAST did not perform as well (55 +/- 32% of control).

**ROC curve performance-** For PSI-BLAST searches, a sequence database is generated, and an all-against-all comparison is made. To effectively identify homologous sequences, the maximum number of iterations is 20. The minimum e-value to include sequences in the construction of a profile for the next iteration is set as 0.0005[2]. The minimum e-value of an alignment to be returned is set as 1,000.

SAM (*S*equence *A*lignment and *M*odeling) [51,52] provides a collection of tools to analyze biological sequences based on linear hidden Markov model. For homology detection, we used SAM-T2K method included in SAM 3.5 package. We first need to generate a HMM model for each 534 query sequence from SABmark twilight zone set. Using *target2k*, each query sequence is searched against the non-redundant protein sequences, NCBI NR database, and we returned a multiple alignment of sequences similar to the query. From the multiple sequence alignment, *w0.5* builds a HMM model for the query. Then, to find homologous sequences of the query sequence, all 533 other sequence are scored based on the model, by *hmmscore*. The Smith-Waterman algorithm is used for the scoring, by default. The 533 sequences are sorted by their E-values. For ROC analysis, we considered first k sequences with the smallest E-values as positives for a query. NCBI NR database, which has 6,419,591 sequences, was downloaded from ftp://ftp.ncbi.nih.gov/blast/db/FASTA/nr.gz (April, 2008). For the NR database searching, *formatdb* and *blastall* in NCBI BLAST 2.2.15 package are used.

For GDDA-BLAST, all protein sequences are encoded as a vector of composite scores (%coverage * %identity * # of hits) for CDD domain profiles. As above, each of the sequences is compared in an all-against-all fashion. To compare encoded sequences, Pearson's correlation coefficient is computed to measure the similarity between two sequences. All values for the Pearson's correlation are accepted into the comparative analysis.

If a sequence pair is actually related or homologous (in the same group) to the query, it is a termed a *true positive*. If a sequence is in the *K* sequences and not related to the query, it is a *false positive*. If a sequence related to the query is not in the *K* sequences, it is a *false negative*. If a sequence not related to the query is not in the *K* sequences, it is a *true negative*. ROC performance analysis for GDDA-BLAST, SAM-T2K, and PSI-BLAST were performed by top-K analysis [53] with largest correlation coefficient for the former or smallest e-value for both SAM-T2K and PSI-BLAST. This sequence is considered as homologous sequence to the query.

**Additional Algorithms Utilized-**
**Domain detection:**

CDD (www.ncbi.nlm.nih.gov/Structure/cdd/wrpsb.cgi)
Pfam (www.sanger.ac.uk/Software/Pfam/)
SMART (http://smart.embl-heidelberg.de/)
Prosite (http://ca.expasy.org/prosite/)
InterProScan (http://www.ebi.ac.uk/InterProScan/)
TMHMM (www.cbs.dtu.dk/services/TMHMM/)

**Unknown Human sequences:** We searched the NCBI protein database for human sequences which contained the key words: unknown and neuronal. Sequences were selected for which ESTs were available and had no conserved domain predictions as measured by rps-BLAST, SMART, pFAM, InterProScan, and Prosite. From these sequences, we randomly chose 3 for careful study using GDDA-BLAST and experiment.

**Lipid-binding Liposomes:** performed as previously described[39]. Briefly, lipid mixtures phosphatidylethanolamine (29.2%), phosphatidylcholine (29.2%), phosphatidylserine (29.2%), and phosphatidylinositol (12.5%) (all in $CHCl_3$) were dried down to form a thin film in a 0.5-ml minifuge tube (Beckmann) and then bath sonicated in 0.2 M sucrose, 20 mM KCl, 20 mM Hepes, pH 7.4, 0.01% azide to yield a 10× dense lipid stock. This was diluted 1:10 in dilution buffer (0.12 M NaCl, 1 mM EGTA, 0.2 mM $CaCl_2$ (free $Ca^{2+}$ concentration of approximately 50 nM), 1.5 mM $MgCl_2$, 1 mM dithiothreitol, 5 mM KCl, 20 mM Hepes, pH 7.4, 1 mg/ml bovine serum albumin) containing 500-1000 ng of recombinant protein. Protein complexes were allowed to form by incubation at 30 °C for 5 min prior to centrifugation (100,000 × *g* for 30 min). After spinning, supernatants were carefully removed and the pellets retrieved by addition of an equal volume of 60 °C SDS sample buffer and subsequent bath sonication. Liposomal assays were visualized via Western analysis. Films were scanned and analyzed using the Bio-rad Gel-dock© system.

**Protein purification-** The Qiagen X-press© protein synthesis *in vitro* translation kit was used as per manufacturer's instructions.

**Antibodies and Reagents-**
Plasmids were from the following sources: HIS Pet28c vectors from Novagen (San Diego, CA). Anti-His antibodies were from Sigma (St. Louis, MO). Lipids were from Avanti (Alabaster, AL). Antibodies were used as per the manufacturer's instructions.